\documentclass[aps,prl,twocolumn,eqsecnum,superscriptaddress,floatfix,10pt]{revtex4-2}
\usepackage{amsmath,amssymb,amsfonts,stmaryrd,wasysym,graphicx,multirow,color,textcomp,subfigure}
\usepackage{url}
\usepackage[colorlinks=true,citecolor=blue,urlcolor=blue]{hyperref}
\usepackage{physics}
\usepackage{chngcntr}
\usepackage{nicefrac}

\usepackage{multirow}
\usepackage{ulem}

\newcommand{\cyan}[1]{{\color{cyan}{\it {#1}}}} 

\def\be{\begin{equation}}
\def\ee{\end{equation}}
\def\bea{\begin{eqnarray}}
\def\eea{\end{eqnarray}}

\counterwithout{equation}{section}

\begin{document}

\title{Eigenstate switching of topologically ordered states \\
using non-Hermitian perturbations}
\author{Cheol Hun Yeom}
\address{Department of Physics, Konkuk University, Seoul 05029, Republic of Korea}
\address{Department of Physics, Hanyang University, Seoul 04763, Republic of Korea}
\author {Beom Hyun Kim}
\email{bomisu@ibs.re.kr}
\affiliation{Center for Theoretical Physics of Complex Systems, Institute for Basic Science, Daejeon 34126, Republic of Korea}
\author{Moon Jip Park}
\email{moonjippark@hanyang.ac.kr}
\address{Department of Physics, Hanyang University, Seoul 04763, Republic of Korea}
\date{\today}

\begin{abstract}
Topologically ordered phases have robust degenerate ground states against the local perturbations, providing a promising platform for fault-tolerant quantum computation. Despite of the non-local feature of the topological order, we find that local non-Hermitian perturbations can induce the transition between the topologically ordered ground states. In this work, we study the toric code in the presence of non-Hermitian perturbations. By controlling the non-Hermiticity, we show that non-orthogonal ground states can exhibit an eigenstate coalescence and have the spectral singularity, known as an exceptional point (EP). We explore the potential of the EPs in the control of topological order. Adiabatic encircling EPs allows for the controlled switching of eigenstates, enabling dynamic manipulation between the ground state degeneracy. Interestingly, we show a property of our scheme that arbitrary strengths of local perturbations can induce the EP and eigenstate switching. Finally, we also show the orientation-dependent behavior of non-adiabatic transitions (NAT) during the dynamic encirclement around an EP. Our work shows that control of the non-Hermiticity can serve as a promising strategy for fault-tolerant quantum information processing.

\end{abstract}

\maketitle

\cyan{Introduction--} The topological quantum computation (TQC) relies on the non-local embedding of information~\cite{Freedman2002,RevModPhys.80.1083,PhysRevLett.97.050401}. By storing the information in a non-local manner, the topologically ordered phase can be a promising platform for robust information storage against undesirable local perturbations~\cite{bravyi2010topological,PhysRevB.41.9377}. The toric code model~\cite{bravyi1998quantum,kitaev2003fault,kitaev2009topological} plays a representative example of a topologically ordered phase, which has been widely accepted and realized in various physical platforms including quantum information~\cite{satzinger2021realizing,PhysRevLett.107.210501} and condensed matter physics~\cite{jiang2012identifying, rodriguez2019identifying, cincio2013characterizing}. Recently, there have been also generalizations in higher dimensions~\cite{castelnovo2008topological,breuckmann2016local,kubica2022single,bravyi2013quantum,bravyi2011topological}. While robust information storage is a prerequisite for any type of TQC, the full information processing demands an effective manipulation scheme of such information.~\cite{PhysRevLett.110.210602,PhysRevB.88.125117,santra2014local} 

Along this line, manipulation and readout schemes for TQC have been suggested in various ways.~\cite{alicea2011non,PhysRevLett.101.260501,vijay2015majorana}. The previous proposals generally assume the unitary time evolution driven by the Hermitian response. However, in an open system, a finite exchange with the reservoir generically gives the gain and loss of energies. In such a case, non-Hermitian spectral singularities, known as exceptional points (EPs), can emerge with the coalescence of the energy eigenstates~\cite{WDHeiss_1990,WDHeiss_2004,chen2020revealing,doi:10.1126/science.abd8872,ding2016emergence,miri2019exceptional,luitz2019exceptional,heiss2012physics}. While the ground state degeneracy is protected by the topological order, we show that the orthogonal property of the eigenstates is not immune against local non-Hermiticity. As such, the eigenstate coalescence in the ground state manifolds can occur by imposing a local non-Hermitian perturbation.

In this work, we propose to utilize the EPs for the manipulation of the topological order. The topologically protected eigenstates can be perfectly switched with adiabatically encircling parturition around EPs. By applying $\mathcal{PT}$-symmetric perturbation, we realize the eigenstate coalescence in the ground state manifolds of the toric code. As mentioned, the adiabatic encircling of the EP allows the eigenstate switching between the topologically degenerate ground states. An advantage of this scheme is that \textit{an arbitrary strength of local perturbations} can induce the EP and, consequently, the perfect eigenstate switching. However, even in the quasi-adiabatic regime, the dynamic effect results in the failure of the adiabatic theorem which gives rise to a time-asymmetric behavior known as the non-adiabatic transition (NAT), which exhibits a dependence on its winding direction while encircling around an EP~\cite{song2021plasmonic, ding2022non, zhang2019distinct, hassan2017dynamically, doppler2016dynamically,liu2021dynamically,zhang2019dynamically,yu2021general,zhang2019dynamically2}.

\begin{figure}[!h]
    \centering
    \includegraphics[width=1.0\linewidth]{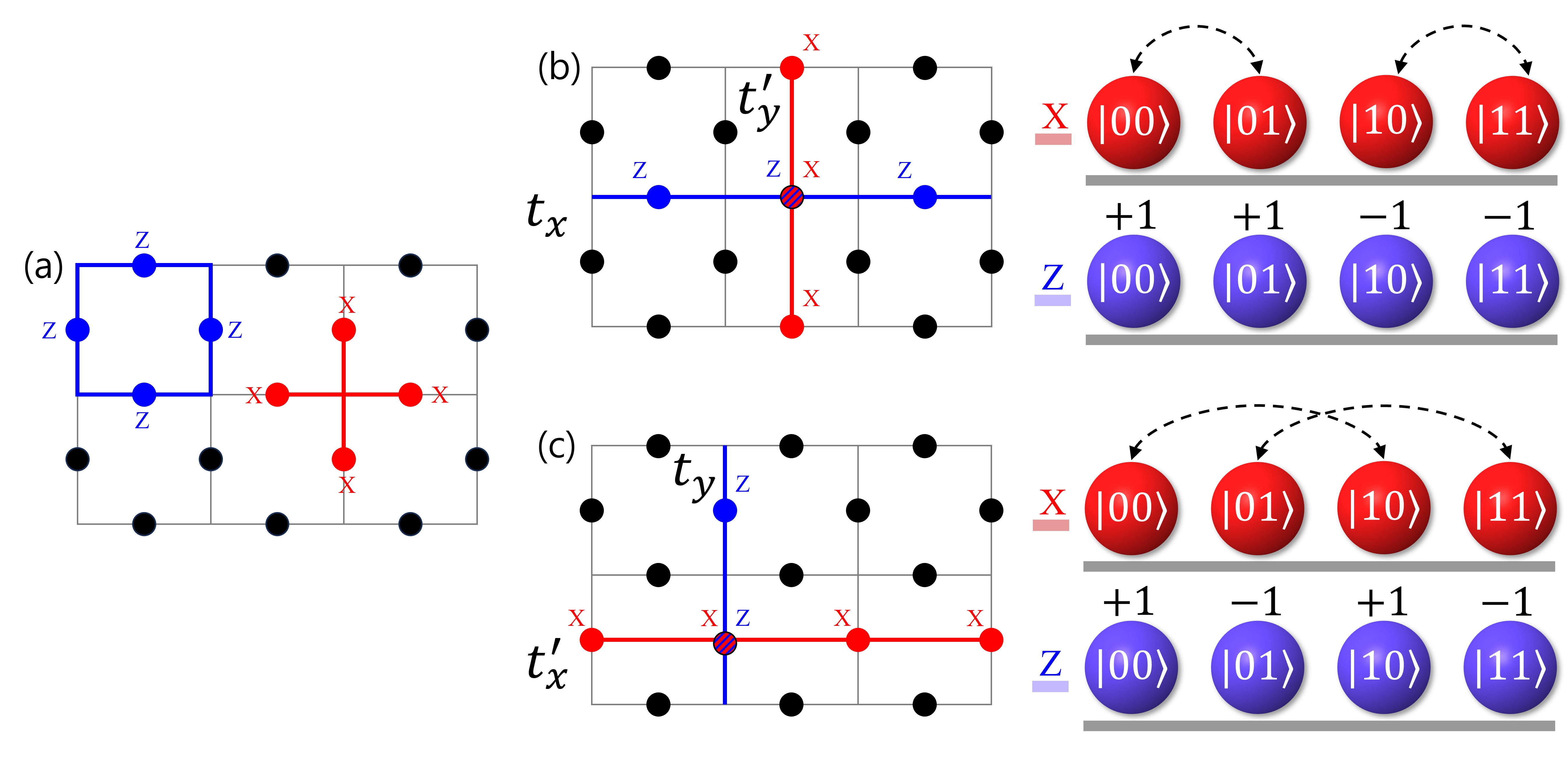}
    \caption{(a) Schematic illustration of vertex and plaquette operators on toric code. (b)-(c) Different perturbations with non-contractible string (NCS) operators. Among the four-fold degenerate ground states, the red NCS operators hybridize between the ground states with distinct parity sectors, while the blue NCS operators energetically split the ground states.}
    \label{fig:1}
\end{figure}

\cyan{Non-Hermitian toric code model--} We begin our discussion by introducing the toric code model~\cite{kitaev2003fault}, where the Hamiltonian can be written as, $H_{\textrm{TC}} =  -\sum_{v} A_v - \sum_{p} B_p$, 
here the vertex(plaquette) operators $A_v$ ($B_p$) indicates the products of the four spin $X$($Z$) operators centered at a vertex $v$(plaquette $p$) [see Fig.~\ref{fig:1}(a)],
$
    A_v = \prod_{i \in v} X_i, \quad B_p = \prod_{i \in p} Z_i.$
There exist the four-fold degenerate ground states $\ket{{m,n}}$, where $m,n\in \{ 0,1 \}$ indicates the parity, that satisfies the condition $A_v \ket{{m,n}} = B_p \ket{{m,n}} = \ket{{m,n}}$ for all vertices $v$ and plaquettes $p$. (See supplementary materials for detailed definitions of ground states). We can define non-contractible strings of Pauli $Z$ or $X$ operators that wind around the dual position of the lattice, $W_{x(y)}=\prod_{i\in t_{x(y)}} Z_i$, $V_{x(y)}=\prod_{i\in t_{x(y)}^{'}} X_i$ [see Fig.~\ref{fig:1}(b)]. $t_x$ ($t_y'$) represents the non-contractible string (NCS) wrapping along $x$$(y)$-direction [See Fig. \ref{fig:1}(b)]. The NCSs map between different ground states among the four-fold topological degeneracy.

While the Hermiticity ensures that four-fold ground states are orthogonal, the introduction of the non-Hermitian perturbations introduces a finite wave function overlap between them. The orthonormality of eigenstates in the ground manifolds is no longer necessary. The eigenstate coalescence, dubbed as EP, can occur. To explicitly see this, we first consider the perturbation with non-Hermitian NCS operators as,
\begin{align} \label{eq:toriccode}
    H = H_{\textrm{TC}} + i\alpha W_x + i\beta V_y.
\end{align}
The coefficients $\alpha$ and $\beta$ are the perturbation strengths of the non-Hermitian string operators.

As we label the basis of ground state degeneracy 
$\ket{\psi}=(\ket{{1,1}},\ket{{1,0}},\ket{{0,1}},\ket{{0,0}})^T$, the matrix form of the perturbation to the ground state sectors can be written as the  effective Hamiltonian as,
\begin{align} \label{effective_H}
    H^{\text{gs}} =\epsilon_0 \textrm{I}-i\begin{pmatrix} 
   \alpha & 0 & \beta & 0  \\
   0 & \alpha & 0 & \beta  \\
   \beta & 0 & -\alpha & 0 \\
   0 & \beta & 0 & -\alpha \\
   \end{pmatrix},
\end{align}
where $\epsilon_0$ denotes the ground state energy of the unperturbed Hamiltonian $H_\text{TC}$. Due to the perturbation, the four-fold ground state degeneracy split into the two-fold degeneracy while the $y$-directional parity, $n$, remains a good quantum number. The split energy eigenvalues are given as, 
$E_{\pm}=\epsilon_0 \pm i\sqrt{\alpha^2+\beta^2}$, which exhibits the typical $\mathcal{PT}$-phase transition when $i\alpha=\pm\beta$. We note that even at the microscopic model, the full Hamiltonian respects $\mathcal{PT}$-symmetry. The $\mathcal{PT}$-symmetry ensures that the energy eigenvalue spectra appear as either a real value ($\mathcal{PT}$-exact, $\alpha^2+\beta^2>0$) or complex conjugate pair ($\mathcal{PT}$-broken, $\alpha^2+\beta^2<0$) \cite{10.1063/1.1418246}. The $\mathcal{PT}$-phase transition is characterized by the emergence of the eigenstate coalescence of the ground state degeneracy. We further confirmed the emergence of the EP using the exact diagonalization of the full lattice Hamiltonian in Eq.~\eqref{eq:toriccode}. We note that there exist different choices of the NCS perturbation that results in the EP as shown in Fig.~\ref{fig:1}(c).

\begin{figure}[t!]
		\centering
		\includegraphics[width=1.0\linewidth]{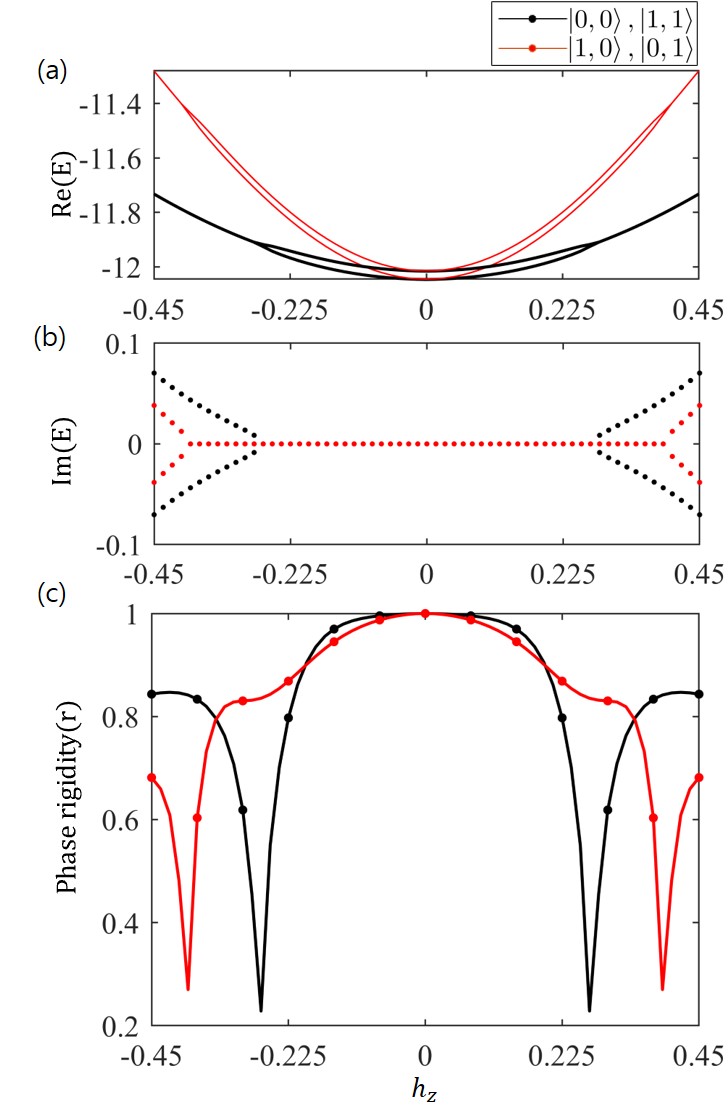}
		\caption{Complex energy spectra of four-fold ground state degeneracy of $H_\textrm{TC}+V$ as a function of the imaginary magnetic field given in Eq.~\eqref{eq:localmagfield} for $L_x=3$ and $L_y=2$. The real and imaginary parts of the energy are indicated in (a) and (b), respectively. (c) Corresponding phase rigidity. The exceptional point arises at different values of $h_z$ with $h_x$ fixed to $0.1$, as confirmed by the vanishing phase rigidity.}
		\label{fig2}
		\centering
\end{figure}

\cyan{Exceptional points under local perturbations--} Although the topologically ordered ground state degeneracy is robust against small local perturbations, the eigenstate orthogonality can be fragile. In principle, local perturbations can also generate the EP between the ground states. To explicitly show this, we perform the exact diagonalization analysis in $L_x \cross L_y$ number of lattice in the presence of the real(imaginary) magnetic field $h_x(h_z)$ along $x$($z$)-direction as, 
\begin{align} \label{eq:localmagfield}
    V =   h_x  \sum_{i\in \textrm{all sites}} X_i +i h_z \sum_{i\in \textrm{all sites}} Z_i.
\end{align}
Fig. \ref{fig2} (a)-(b) shows the complex energy spectra in the odd number of system sizes as a function of the magnetic field. Unlike the case of the NCS perturbation, the ground state energy corrections are suppressed in the thermodynamic limit. Nevertheless, we observe the emergence of the EP and the $\mathcal{PT}$-phase transition accompanied by vanishing phase rigidity, $r$, which measures the eigenstate coalescence as~\cite{PhysRevE.74.056204},
\bea
r_{m,n} = \abs{\frac{\braket{(m,n)_L}{(m,n)_R}}{\braket{(m,n)_R}{(m,n)_R}}},
\eea
where $|(m,n)_{L(R)}\rangle$ is the left(right) ground state adiabatically deformed from the original eigenstate with the definite parity. The emergence of the EP persists even in large system sizes since the eigenstate coalescence only depends on the relative strength of the $x-$ and $z-$directional imaginary magnetic field. The critical ratio of the magnetic field $h_x/h_z$ where the EP arises depends on the system size $L_x$ and $L_z$. 

Therefore, any arbitrary strength of the non-Hermitian perturbation can cause the $\mathcal{PT}$-phase transition.  As the system size of the toric code increases, the splitting of the ground state energies is exponentially suppressed due to the topological protection~\cite{PhysRevLett.106.107203}. Nevertheless, the emergence of the EP is robust, and it only depends on the ratio of $h_z/h_x$ regardless of the system size. The EP occurs since there exist non-zero matrix elements between the ground states in high-order perturbations. As we discuss more in detail below, it is important to point out that the suppressed splitting in the thermodynamic limit can practically limit the lower bound of the quasi-adiabatic driving timescale since the long-timescale driving smaller than the split energy can lead to different non-adiabatic behavior compared to short-timescale dynamic driving.

It is important to note that, when $L_x$ and $L_y$ are both even numbers, the generation of the EP between the ground states is not guaranteed due to the absence of the effective $\mathcal{PT}$ symmetry~\cite{shackleton2020protection}. We can define the pseudo-Hermitian condition, $\eta(H_{\mathrm{TC}}+V)\eta^{\dagger}=(H_{\mathrm{TC}}+V)^\dagger$, where $\eta=\prod_{i}Y_i$ is a product of $Y$ operators at all sites. The pseudo-Hermiticity ensures that the eigenvalues come as the complex conjugate pairs. When $L_x$ and $L_y$ are both even integers, all the ground states are characterized by the same eigenvalues of $\eta$ since the NCS operator commutes with $\eta$. However, when the number of lattices in either direction is odd, the two of the ground states have the eigenvalue $\eta=+1$ while the others have $\eta=-1$. When the eigenvalue of the pseudo-Hermitian operator $\eta$ is the same, the complex energy must stay in the real axis, which forbids the possibility of the emergence of the EP~\cite{krein1950generalization,Gel'fand1955}. 



\cyan{Adiabatic encircling of the exceptional point--} By adiabatically encircling the EP, the eigenstate switching effect can be utilized for possible manipulation of the ground state. The EP is the topological branch point where the complex energy spectra of different ground states meet. The adiabatic encircling of an EP leads to the interchange of eigenstates between the two ground states. By encircling the EP twice, the state returns to itself with the accrued Berry phase of $\pi$. If the system is completely adiabatic, one can observe the instantaneous eigenstate flip, such that the lower state ends up with the upper state or vice versa at the end of the adiabatic cyclic driving. When the EP occurs between more than two states, the eigenstate switch effect is described by the relevant permutation group \cite{PhysRevA.106.012218}. We note that the state-exchange mechanism does not depend on its winding direction in a completely adiabatic condition.

\begin{figure*}[htbp]
		\centering
		\includegraphics[width=1.0\linewidth]{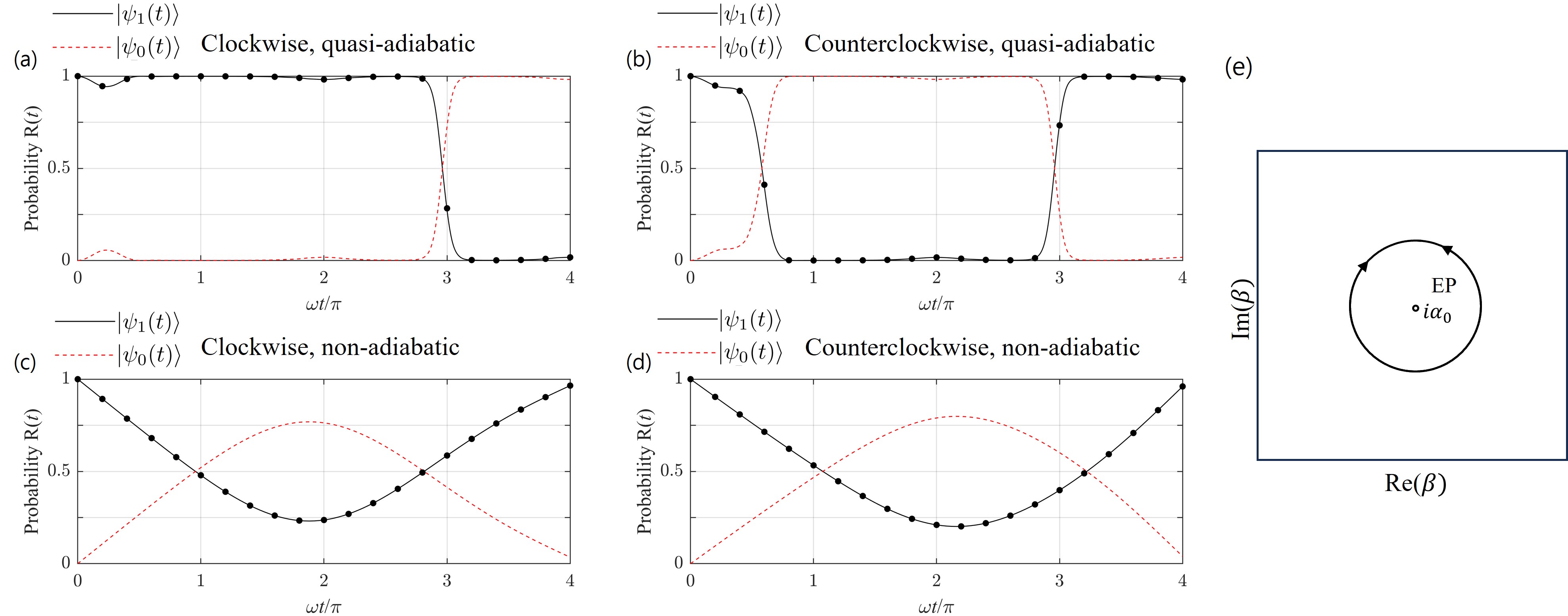}
		\caption{Dynamic driving of the entangled state $\ket{\psi(t)}$. (a)-(b) Time evolution under the quasi-adiabatic limit; clockwise and counterclockwise driving exhibit different numbers of non-adiabatic transitions in two cycles with a period $T$. (c)-(d) While driving under the non-adiabatic limit, the fast-timescale effect dominates over non-unitary couplings, which induces the conventional unavoidable state flip. In (a-d) $\ket{\psi_{1(0)}(t)}$ denotes the projection of the state $\ket{\psi(t)}$ to $m=1(0)$ parity sector, and $\omega=2\pi/T$. The red dotted lines indicate the failure rate of the perfect eigenstate switching. (e) The schematic figure for encircling an EP.}
		\label{fig3}
		\centering
\end{figure*}

\cyan{Dynamic encircling of the exceptional point--} In contrast to the adiabatic encircling, when the timescale of the encircling ($T$) is comparable to the energy splitting ($\abs{\Delta E}$) of the ground state degeneracy, $T\sim \abs{\Delta E}$, the non-adiabatic behavior is evidently observed in the short-timescale regime. The non-adiabatic behavior is manifested as a non-negligible failure rate in the eigenstate switching [Fig.~\ref{fig3}(c,d)]. Furthermore, non-Hermitian dynamic driving within the finite time scale always gives rise to the non-zero failure rate in the eigenstate switching~\cite{berry2011slow}. Even in long-timescale dynamic driving, depending on the encircling direction, an asymmetric unavoidable state flip survives, known as a non-adiabatic transition (NAT)~\cite{PhysRevA.92.052124,uzdin2011observability}. To explicitly observe the NAT, we solve the time-dependent Schr\"{o}dinger equation of the NCS perturbation model (Eq.~\eqref{eq:toriccode}) with the initially entangled state $\ket{\psi(t=0)}=a\ket{1,1}+b\ket{1,0}$, where $|a|^2+|b|^2=1$ as,
\begin{align} \label{evolution}
    &i\frac{d}{dt}\ket{\psi(t)} = H(t)\ket{\psi(t)}, \\
    &H(t) = H_{\textrm{TC}}+\alpha(t)W_x+\beta(t)V_y,
\end{align}
where $\alpha(t)=\alpha_0$ and $\beta(t) = i\alpha_0(1-e^{\pm 2\pi i t/T})$ describe the time-dependent encircling around the EP, where the sign in the exponent $\beta(t)$ indicates the direction of encircling: the counterclockwise$(+)$ and clockwise$(-)$, respectively, as illustrated in Fig.~\ref{fig3}(e). Initially, the system starts with the $\mathcal{PT}$-exact Hamiltonian $(\beta=0)$, where the four basis states are mutually orthogonal. During the driving, the instantaneous four-fold ground states are always split into two-fold degeneracy with the different $x$-directional parity sector $m$. The adiabatic encircling of the EP switches the eigenstates between the distinct $x$-directional parity sector $m$. To quantify the NAT, we introduce the probability amplitude $R(t)$ of the state staying $m=1$ parity sector as,
\bea
R(t) = \frac{\sum_{n=0,1}\abs{\braket{m=1,n_{(t)}}{\psi(t)}}^2}{\sum_{m,n=0,1}\abs{\braket{m,n_{(t)}}{\psi(t)}}^2}, 
\eea
where $\bra{m,n_{(t)}}$ denotes the instantaneous left ground states of $H(t)$ at the time $t$. 

The small probability amplitude, $R(t)$, indicates the occurrence of the unavoidable state flip, where the time-evolved state undergoes the transition to the $m=0$ parity sector, which deviates from the initial $m=1$ parity sector. In the non-adiabatic driving [Fig. \ref{fig3}(c),(d)], the gradual and smooth evolution of $R(t)$ is similar to the phenomenology of the Landau-Zener transition~\cite{landau1932theorie,zener1932non}. On the other hand, the abrupt jump of NAT in the adiabatic regime [Fig. \ref{fig3}(a),(b)] is distinctly observed from the conventional behavior of the Hermitian Landau-Zener transition. During the process of the time-evolution, the instantaneous $\mathcal{PT}$-broken Hamiltonian ($\beta\neq 0$) contributes asymmetry in the population as one of the parity sectors has exponentially growing amplitude while the amplitude of the other is exponentially decaying. The resulting asymmetry in the population is proportional to the period $T$ of the driving. In the adiabatic regime, the asymmetry in the population is increased. In this case, the dependence of the winding direction of the driving becomes crucial in the behavior of the probability amplitude $R(t)$. For instance, if the system is driven in a clockwise direction [Fig. \ref{fig3}(a)], one complete cycle of driving around the exceptional point ($\omega t /\pi = 2$) can perfectly switch the initial ground state $\ket{\psi(t=0)}$.~[$\ket{m,n}\rightarrow\ket{m+1,n}(\text{mod}~2)$]. On the other hand, a complete cycle in the counter-clockwise driving, as shown in Fig. \ref{fig3}(b), shows the zero-switching rate ($R(t)=0$).

\cyan{Discussions--} 
Let us briefly consider the dynamic driving under the local perturbation. Unlike the case of the non-local string perturbation, introducing a finite magnetic field can induce the phase transition from the topologically ordered phase to the paramagnetic phase~\cite{PhysRevB.100.125159}. Nevertheless, as previously noted, a significant benefit of applying local perturbation lies in the fact that the generation of an EP is determined solely by the ratio of two non-commuting operators. This suggests that EPs can be generated even in the large system size, within a regime of infinitesimally small perturbations where the topological order remains intact. Consequently, this allows for the process of information manipulation to be carried out effectively under the influence of local perturbations.

In conclusion, we have demonstrated the generation of the EP  between topologically ordered ground states of the toric code by introducing non-Hermitian perturbations. While the topological degeneracy is protected against the local perturbation, the eigenstate coalescence can be induced by any infinitesimal strength of the local perturbations. We have argued that the switching effect between the ground states can be performed by dynamically encircling around an EP. In practice, we expect the directional-asymmetric non-adiabatic transition in the slow-timescale dynamics, which gives rise to a non-negligible failure rate of the eigenstate switching effect. On the other hand, the controllable non-Hermiticity by using gain-and-loss has been recently realized in programmable quantum computing platforms~\cite{shen2023observation} as well as the Hermitian toric code model~\cite{satzinger2021realizing}. One effective method to realize this is to generate an EP and the associated eigenstate switching effect by using a local field. A notable property of non-Hermitian local perturbations is that they allow us to make an EP with an infinitesimally small strength of perturbations, preserving the topological order of the ground states while achieving the switching effect. Our proposal provides a fundamentally advantageous approach for controlling topological order utilizing non-Hermitian topological physics.

\begin{acknowledgments}
\section{Acknowledgement}
M.J.P. thanks Dung Xuan Nguyen and Yong Baek Kim for fruitful discussions. This work was supported by the National Research Foundation of Korea
(NRF) grant funded by the Korea government (MSIT) (Grants No. RS-2023-00252085 and No. RS-2023-00218998).  B.H.K. was supported by the Institute for Basic Science in Korea (IBS-R024-D1).
\end{acknowledgments}

\bibliography{reference_main2}
\newpage
\pagebreak
\onecolumngrid
\renewcommand{\thefigure}{S\arabic{figure}}
\setcounter{equation}{0}

\section*{Supplementary Materials}
\subsection{Derivation of the effective Hamiltonian for non-Hermitian toric code}
We define the Wilson loop composed of $Z$-strings such that $W_x=\prod_{i\in t_x} Z_i$ and $ W_y=\prod_{i\in t_y} Z_i$, where $t_{x,y}$ indicates the spins located at the position of $x$($y$)-directional non-contractible strings. In addition, we can define the dual Wilson loop composed of $X$-strings, $V_x=\prod_{i\in t_x^{'}} X_i$ and $V_y=\prod_{i\in t_y^{'}} X_i$. [See Fig.~\ref{fig:1}(b,c) in the main text.] The four-fold ground state degeneracy of the topological order can be labeled by the eigenvalues of $W_{x,y}$, while the dual Wilson loop $V_{x,y}$ maps between the different ground states as,
\bea
\label{Eqs:VW}
& W_x\ket{m,n}=(-1)^m\ket{m,n}, \\
& W_y\ket{m,n}=(-1)^n\ket{m,n}, \\
& \ket{m,n}=(-1)^{m+n}(V_x)^m(V_y)^n\ket{0,0},\\
& \ket{0,0} = C \prod_v(1+A_v)\ket{\text{vac}},
\eea
where $C$ is the normalization coefficient and the state $\ket{\text{vac}}$ is the state in which all spins are up, $\ket{\text{vacuum}}=\ket{\uparrow\uparrow\uparrow\uparrow\cdots}$.
As we identify each ground state with a four-dimensional vector as,
\bea
&\ket{1,1} \rightarrow (1,0,0,0)^T\\
&\ket{1,0} \rightarrow (0,1,0,0)^T\\
&\ket{0,1} \rightarrow (0,0,1,0)^T\\
&\ket{0,0} \rightarrow (0,0,0,1)^T.
\eea
we can derive the matrix form [$(\mathcal{M}_{ij})=\bra{i}\mathcal{M}\ket{j}$] of the Wilson loop operators using Eq. \eqref{Eqs:VW} as following,

\bea
& W_x = \begin{pmatrix} 
   -1 & 0 & 0 & 0  \\
   0 & -1 & 0 & 0  \\
   0 & 0 & 1 & 0 \\
   0 & 0 & 0 & 1 \\
   \end{pmatrix}, 
& W_y = \begin{pmatrix} 
   -1 & 0 & 0 & 0  \\
   0 & 1 & 0 & 0  \\
   0 & 0 & -1 & 0 \\
   0 & 0 & 0 & 1 \\
   \end{pmatrix},\\
& V_x = \begin{pmatrix} 
   0 & -1 & 0 & 0  \\
   -1 & 0 & 0 & 0  \\
   0 & 0 & 0 & -1 \\
   0 & 0 & -1 & 0 \\
   \end{pmatrix}, 
& V_y = \begin{pmatrix} 
   0 & 0 & -1 & 0  \\
   0 & 0 & 0 & -1  \\
   -1 & 0 & 0 & 0 \\
   0 & -1 & 0 & 0 \\
   \end{pmatrix}.
\eea

\subsection{Numerical Calculation of Berry Phase}
In the adiabatic evolution, the complex Berry phase $\gamma$ is defined as
\begin{align}
    \gamma = i \oint_{\mathcal{C}_\beta} \frac{\braket{\phi(t)}{\partial_t \psi(t)}}{\braket{\phi(t)}{\psi(t)}},
\end{align}
where ${\mathcal{C}_\beta}:\mathbb{R} \rightarrow \mathcal{M} $ is a loop where $\mathcal{M}$ denotes the Riemann surface of the energy, and $\bra{\phi(t)}(\ket{\psi(t)})$ denotes the left(right) eigenstates of the Hamiltonian $H(t)$. The loop ${\mathcal{C}_\beta}$ is parameterized by 
\begin{align}
    \mathcal{C}_\beta : t \in [0,4\pi] \mapsto E(\beta(t)), \quad \beta(t) = i\alpha_0 (1-e^{iwt}),
\end{align}
where $E(t)$ is the instantaneous eigenvalue of $H(t)$ and $\beta(t)=\alpha_0$ indicates an exceptional point.

For the discrete adiabatic time evolution with large N steps, we can approximate the integral $\gamma$ as a finite sum of the phase element. Then, the Berry phase is given by
\begin{align}
    \gamma & = -\text{Im}\log{\left( \prod_{i=0}^{N} \braket{\phi(t_{i+1})}{\psi(t_i)} \right)} \\
    & = -\text{Im}\log{\left( \braket{\phi(t_{N+1})}{\psi(t_{N})}\braket{\phi(t_{N})}{\psi(t_{N-1})}\braket{\phi(t_{N-1})}{\psi(t_{N-2})} \cdots \braket{\phi(t_{2})}{\psi(t_{1})}\braket{\phi(t_{1})}{\psi(t_{0})} \right)},
\end{align}
where $t_0 = 0$, $t_N = T$, $t_{N+1} = t_{0}$ and $\Delta t = T/N$ with $T=4\pi/\omega$. To realize the smooth growth of the Berry phase, we give a smooth complex gauge to each $\bra{\phi(t_i)}$ and $\ket{\psi(t_i)}$ by following. For the set $\{\bra{\phi(t_i)}\}$,$\{\ket{\psi(t_i)}\}$ of the instantaneous left(right) eigenstates, we first give a parallel-transport gauge condition, $\braket{\phi(t_{i+1})}{\psi(t_i)}>0$ by the transformation
\begin{align}
    &\ket{\psi(t_{i+1})} \mapsto e^{i(\text{Im}\log{\braket{\phi(t_{i+1})}{\psi(t_i)}})}\ket{\psi(t_{i+1})} \\
    &\bra{\phi(t_{i+1})} \mapsto \bra{\phi(t_{i+1})}e^{-i(\text{Im}\log{\braket{\phi(t_{i+1})}{\psi(t_i)}})},
\end{align}
for $i=1,2,\cdots,N$. In the parallel transport gauge, the Berry phase is given by
\begin{align}
    \gamma & = -\text{Im}\log{\left( \braket{\phi(t_{N+1})}{\psi(t_{N})}\braket{\phi(t_{N})}{\psi(t_{N-1})}\braket{\phi(t_{N-1})}{\psi(t_{N-2})} \cdots \braket{\phi(t_{2})}{\psi(t_{1})}\braket{\phi(t_{1})}{\psi(t_{0})} \right)} \\
    & = -\text{Im}\log{\braket{\phi(t_{N+1})}{\psi(t_1)}}.
\end{align}
In the above calculation, the only phase difference at $t_0$ and $t_{N+1}$ gives the Berry phase, which breaks the periodicity of the states. Thus we allocate the additional phase to the states by
\begin{align}
    &\ket{\psi(t_{i+1})} \mapsto e^{i \gamma t_{i+1} / \omega T}\ket{\psi(t_{i+1})} \\
    &\bra{\phi(t_{i+1})} \mapsto \bra{\phi(t_{i+1})}e^{-i \gamma t_{i+1} / \omega T} .
\end{align}
Then every inner product $\braket{\phi(t_{i+1})}{\psi(t_i)}$ has equal phase amount of $\pi/N$, implying that the total phase amount equals to $\pi$ after calculating $N$ steps.
By adopting this choice of gauge, we have calculated the quantized Berry phase of $\pi$ after winding two full cycles around an EP.
\end{document}